# Ultrafast excitation of Bloch plasmon polaritons in hyperbolic metamaterials with an extreme ultra-violet transient grating

Tlek Tapani[1¶], Hannes Kempf[1¶], Matteo Pancaldi[2¶*], Laura Foglia[2], Emanuele Pedersoli[2], Roberta Totani[2], Adriana Valerio[2], Riccardo Mincigrucci[2], Ivaylo Nikolov[2], Miltcho B. Danailov[2], Aitor De Andrés[1], Roman Krahne[3], Paolo Vavassori[4,5], Filippo Bencivenga[2], Flavio Capotondi[2], Denis Garoli[3,6], and Nicolò Maccaferri[1,7*]

[1]Ultrafast Nanoscience Group, Department of Physics, Umeå University, Linneaus väg 24, 90187 Umeå, Sweden
[2]Elettra Sincrotrone Trieste, Strada Statale 14, 34149 Basovizza, TS, Italy
[3]Optoelectronics Group, Instituto Italiano di Tecnologia, Via Morego 30, 16163 Genova, Italy
[4]CIC nanoGUNE BRTA, Tolosa Hiribidea 76, 20018 Donostia-San Sebastián, Spain
[5]Ikerbasque, Basque Foundation for Science, Plaza Euskadi 5, 48009 Bilbao, Spain
[6]Dipartimento di Scienze e Metodi dell'Ingegneria, Università degli Studi di Modena e Reggio Emilia, Via Amendola 2, 14142 Reggio Emilia, Italy
[7]Wallenberg Initiative Materials Science for Sustainability, Department of Physics, Umeå University, Linneaus väg 24, 90187 Umeå, Sweden
¶Contributed equally
*matteo.pancaldi@elettra.eu
*nicolo.maccaferri@umu.se

**Manipulating materials properties with light drives advances in materials science and photonics. Hyperbolic metamaterials[1] are promising candidates as next-generation quantum optical media[2]. They support Bloch plasmon polaritons[3], which are characterized by potentially infinite wave-vectors and long lifetimes, but cannot be excited through direct light illumination due to momentum mismatch. Here, we experimentally show that a transient grating, formed via interference of fully coherent seeded free-electron laser pulses in a thin insulator film, enables the excitation of Bloch plasmon polaritons in an underlying hyperbolic metamaterial. Finite element simulations confirm the role of the transient grating in facilitating phase-matching and mode excitation. Our findings demonstrate a route to spatiotemporally excite Bloch plasmon polaritons modes, offering an alternative to permanently nanostructured gratings and potentially enabling ultrafast control of optical modes excitation.**

In the past years, hyperbolic metamaterials (HMMs) have attracted great interest due to their optical properties[1], such as negative refraction[4], optical cloaking[5] and tunable absorption and scattering[6], enabling enhanced nonreciprocal light propagation[7], strong optical nonlinearities[8,9], extreme biosensing[10] and super resolution imaging[11]. They display hyperbolic iso-frequency surfaces, meaning that light travelling within these materials can have, in principle, infinite wave-vector[1], as well as reach superluminal speed[12]. From a practical point of view, it is possible to engineer hyperbolic behavior using periodic stacks of metallic and dielectric layers, supporting propagating modes called Bloch plasmon polaritons (BPPs)[3]. These modes exhibit longer lifetimes and propagation distances compared to other optical modes supported by conventional optical materials, making such HMMs good candidates for next-generation quantum optical media[2]. Since BPPs cannot be excited with direct light illumination due to momentum mismatch, they cannot be observed in the far-field, for instance by measuring their reflectance (Fig. 1a-c). However, they can be excited by using patterned surfaces fabricated by, e.g., electron beam lithography[13]. An open question is whether we can excite BPPs by patterning the surface of these metamaterials in time, exploiting the coherent response to an ultrafast excitation to activate these modes.

Periodic temporal modulation can be accessed through laser beam interference[14] to dynamically induce transient gratings (TGs) within homogeneous films[15]. For instance, theoretical predictions showed that TGs consisting of non-equilibrium carrier dynamics-induced permittivity modulation in a semiconductor thin film can provide the required momentum to excite surface plasmon polaritons[16], and recent experiments demonstrated that TGs can act as a symmetry-breaking mechanism to excite photonic quasi-bound states in the continuum[17].

Here, we use an extreme ultraviolet (EUV) TG to photo-induce a spatial modulation of the permittivity in an initially homogeneous electrical insulator, namely a 30 nm thick $Al_2O_3$ thin film. Two intersecting femtosecond EUV pulses produced by a fully coherent seeded free-electron laser (FEL) source create a spatially periodic absorption pattern through interference (Fig. 1d). This process leads to a spatially periodic generation of non-equilibrium carriers, resulting in a modulation of the thin film permittivity. Notably, $Al_2O_3$ is a wide bandgap material, which makes the use of EUV radiation particularly effective for inducing non-equilibrium carrier dynamics and excite valence and core electrons on a time scale below 1 ps[15,16]. In this short temporal window, a BPP can be excited in the HMM with another (probe) pulse (Fig. 1d), as the EUV TG provides the necessary phase-matching condition (Fig. 1e). This manifests as a minimum in far field reflectance measurements (Fig. 1f), as previously shown with static patterned gratings[13]. To support our findings, we simulate the experiment using the finite element method (FEM) and taking into consideration the local spatial variation of the $Al_2O_3$ thin film permittivity. Also, we perform benchmark experiments on the bare $Al_2O_3$ thin film, not coupled to any HMM structure, to provide additional evidence that the observed spectral feature can be ascribed to the excitation of a BPP mode in the HMM by using an EUV TG.

For the experimental investigation, a HMM structure made of alternating layers of [Au (15 nm)/$Al_2O_3$ (30 nm)]x8, was synthetized on a fused silica substrate (for more details see Methods). An extra layer of $Al_2O_3$ (30 nm) was grown on the top Au layer (Fig. 1a). A reference sample with 30 nm of $Al_2O_3$ on top of fused silica was also prepared. A representative scanning electron microscopy image of the multilayered structure is reported in Extended Data Fig. 1.

On this sample, we performed pump-probe measurements recording the transient reflectance $\Delta R/R$ of the HMM (Fig. 2a and b). For more details about the measurements protocol and the experimental setup see Methods and Extended Data Fig. 2. When the EUV beams interact with the sample, they interfere in the $Al_2O_3$ layer generating a TG, whose periodicity is determined by their wavelength and crossing angle. In our case, the period of the TG was set to 383 nm (wavelength: 22.7 nm, crossing angle: 3.4 deg; see Methods for more details). At a pump-probe delay of 0.1 ps, we observe the appearance of a dip in the $\Delta R/R$ signal around 1230 nm (Fig. 2a, red curve).

At this time delay, the non-equilibrium carriers provide the strongest contribution to permittivity modulation[15,16], and the effect due to the TG is at its maximum amplitude (Fig. 2b, red line). We then repeated the measurement with the probe pulse arriving at a time delay of 2 ps from the coherent creation of the TG, such that the permittivity modulation is expected to be less effective due to carrier diffusion[16]. The result (Fig. 2a, yellow curve) shows that the $\Delta R/R$ signal does not present the dip at 1230 nm anymore, but it is rather featureless. This is consistent with the picture that after 1 ps there is not enough permittivity contrast due to carriers diffusion[18,16], and that the excitation of a BPP using a TG is mainly due to the coherent excitation of the latter.

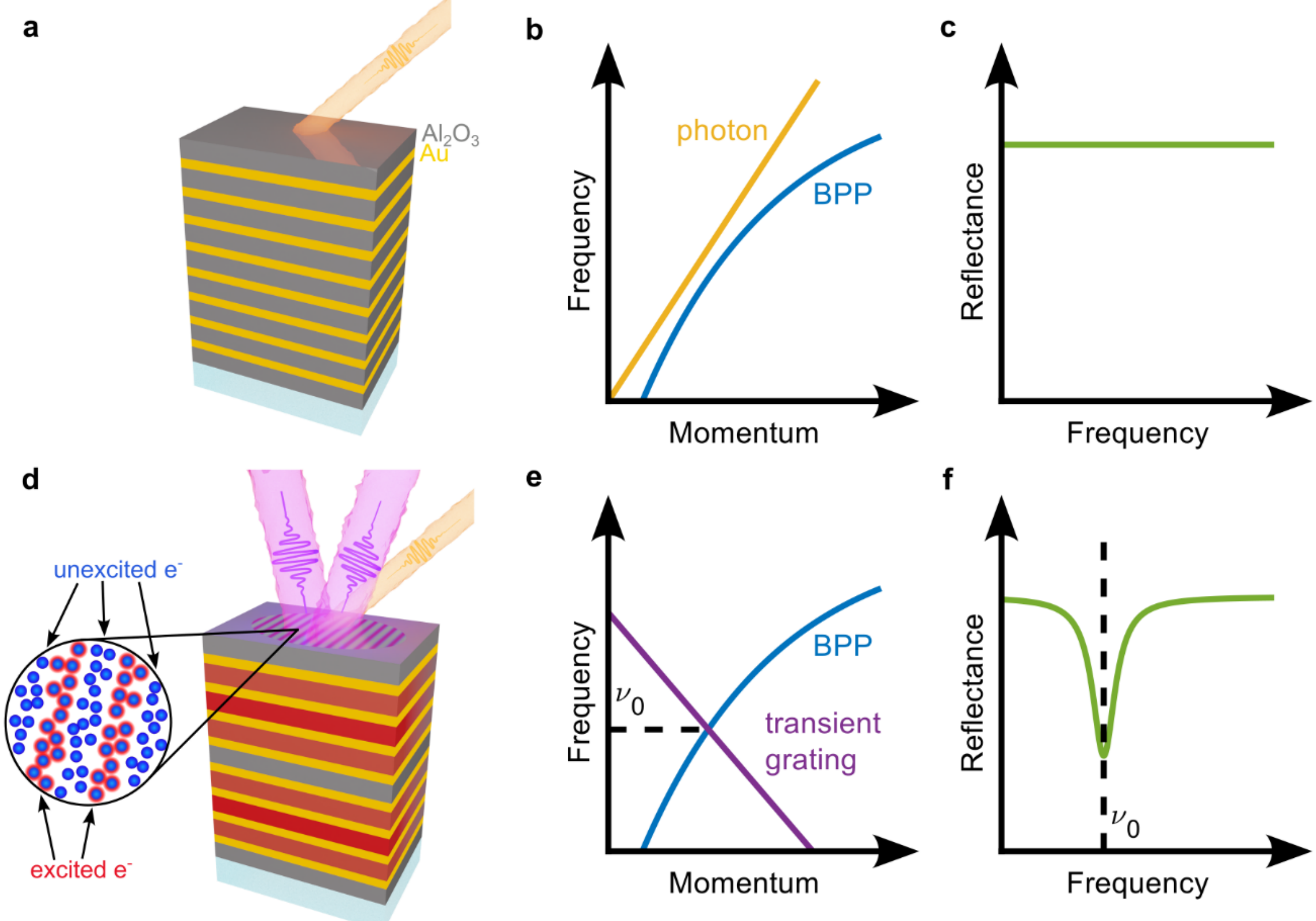

**Fig. 1. Transient grating-coupled hyperbolic metamaterials and Bloch plasmon polariton excitation.** (a) Schematic of the Au/$Al_2O_3$ multilayer structure with a homogeneous $Al_2O_3$ thin film on top. In the absence of a grating on top of this structure or any permittivity modulation, direct illumination does not provide sufficient in-plane momentum to excite a BPP mode. (b) Dispersion diagram illustrating the momentum mismatch between free-space photons and a BPP mode supported by the multilayer structure. (c) Corresponding reflectance spectrum, showing no resonant feature associated with BPP excitation. (d) Two intersecting EUV femtosecond pulses generate a TG in the $Al_2O_3$ thin film through a spatially periodic excitation pattern, producing a periodic modulation of the thin film permittivity. (e) The TG supplies the additional momentum required for phase matching, enabling excitation of the BPP mode at frequency $\nu_0$. (f) As a result, a resonance appears in the reflectance spectrum at $\nu_0$, indicating transient grating enabled excitation of a BPP mode.

As a control experiment, we used one of the two EUV beams acting as the pump, at twice the fluence used when both beams are activated to generate the TG modulation (Fig. 2a, blue curve). This ensures a fair comparison, as the same average energy is deposited on the sample in both cases while eliminating the spatial modulation of the permittivity in the $Al_2O_3$ layer. The absence of BPP features in the $\Delta R/R$ signal over the whole range of the probed wavelengths indicates that the EUV excitation by itself is not able to provide the momentum to excite BPPs with the probe light. This is expected because, in the absence of a TG modulating the permittivity of the $Al_2O_3$ layer, the probe light is reflected by the multilayer structure without the spectral features associated with grating-assisted coupling, as previously observed for HMM without a grating on top[13] (see also Fig. 1a-c, where only the probe light is shown; in this case, adding a EUV pump does not change the result, since the momentum provided in not enough to excite BPPs).
As additional control, we also performed the experiment with TG on the reference $Al_2O_3$ layer without multilayer (see Extended Data Fig. 3), where we did not observe any mode in the spectral range of interest. This further corroborates that the dip in the reflectance at 1230 nm is connected to the excitation of a BPP in the HMM structure via the excitation of a TG.

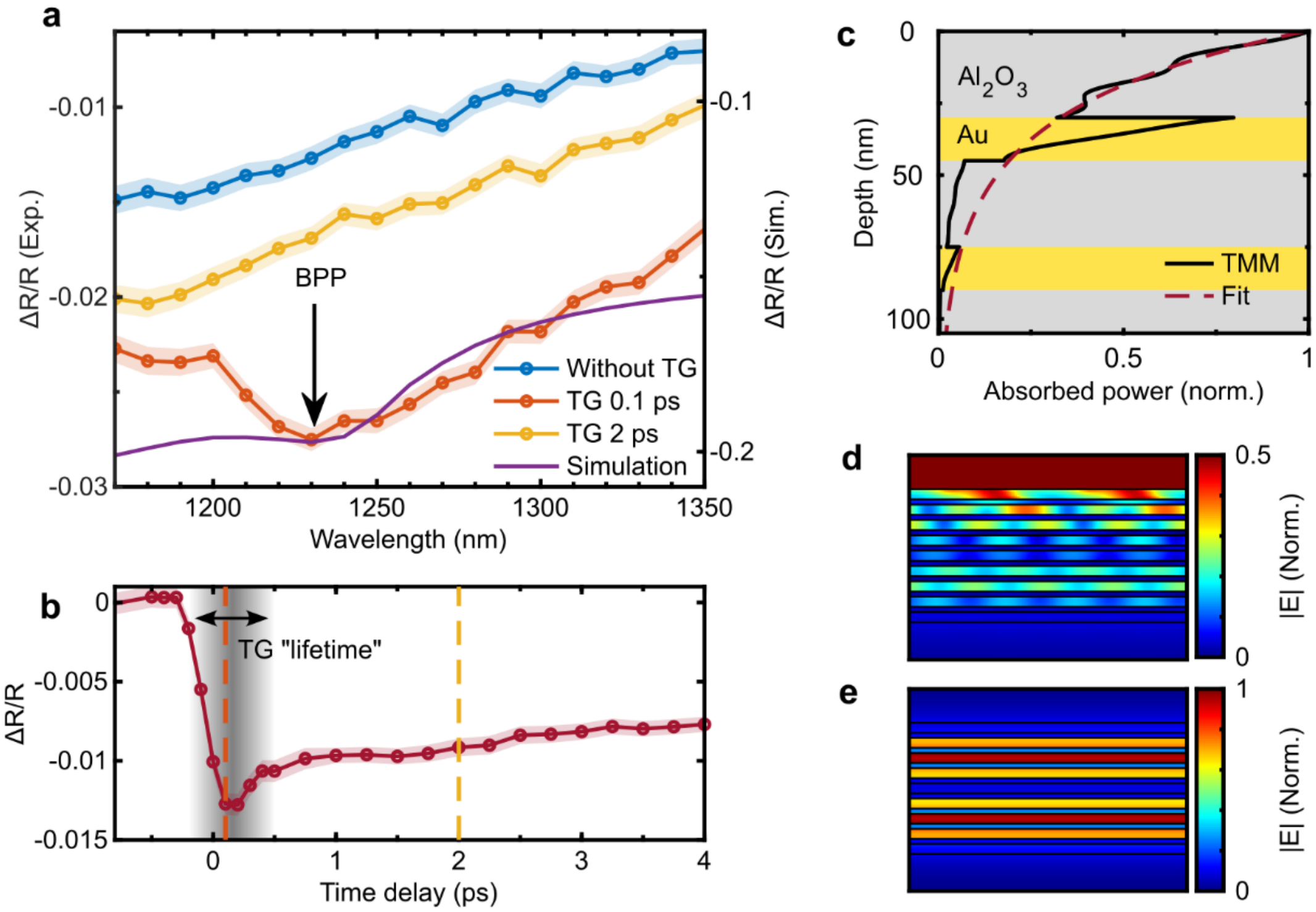

**Fig. 2. Experimental observation of BPP excitation via TG and numerical simulations.** (a) Transient reflectance ΔR/R for three consecutive experiments where we first created the transient grating (two pump pulses interfering at the sample) at different time delays: 0.1 ps (red curve) and 2 ps (yellow curve). The third blue curve is the control experiment, where we did not create a TG (one pump pulse, double fluence, at 0.1 ps time delay). The violet line is the simulated response for a TG period of 383 nm. The black arrow marks the BPP mode at 1230 nm. (b) Transient reflectance ΔR/R as a function of time delay between the two pump (22.7 nm) and the probe (1250 nm) pulses; vertical lines represent chosen time delays: 0.1 ps and 2 ps in (a). (c) TMM calculation (black solid line) of absorbed power as a function of sample depth for p-polarized EUV light, compared with a simple exponential fit (red dashed line) showing a penetration depth of approx. 30 nm. (d) Near-field profile of the excited BPP mode in presence of the TG. (e) Eigenmode of the BPP within the multilayer.

We further elucidated the excitation mechanism by numerically simulating the effect of the TG on the multilayer system using the transfer-matrix method (TMM) implemented in NTMpy[19] and the FEM approach implemented in COMSOL Multiphysics[20]. In Fig. 2c we report the TMM calculation (black solid line) of the absorbed power as a function of the system depth considering a p-polarized beam imping on the sample with an angle of 1.7 deg and a wavelength of 22.7 nm. To give an estimate of the involved sample depth, we compare this calculation with a simple exponential fit (red dashed line) showing a characteristic absorption penetration length of approx. 30 nm. Nonetheless, absorption peaks within the first two [Au (15

nm)/$Al_2O_3$ (30 nm)] bilayers suggest that the generated TG extends beyond the the first $Al_2O_3$ layer into the HMM multilayer structure. As detailed in the Methods section, we used the calculated absorption for modelling the effect of the TG in the FEM simulations: while the absorbed power profile describes the spatial evolution of the grating in the out-of-plane direction, we considered a cosinusoidal modulation of the refractive index along the direction of the grating wavevector, as expected from the interference of the two pump beams. Having defined the shape of our TG, we then compared the simulated ΔR/R signal for a grating period of 383 nm (violet line) with the experimental curve shown in Fig. 2a (orange curve). The arrow marks the experimental BPP mode at 1230 nm. The corresponding near-field distributions of the excited BPP mode for a grating period of 383 nm is shown in Fig. 2d. Fig. 2e shows the distribution of the electric field amplitude of the corresponding eigenmode in the unpatterned HMM obtained via FEM mode analysis. Notably, the field profile in Fig. 2d closely matches the field distribution of the eigenmode of the bare HMM[13], confirming that in our experiments the generation of a TG allows the excitation of BPPs within the multilayer structure.
The optimization of the experimental parameters and the acquisition of the large set of data required for a full investigation summarized in Fig. 2a, entailed a prolonged exposure of a specific portion of the sample surface to the EUV pump. As a consequence, cumulative surface damage could be expected. We therefore repeated a measurement at a time delay of 0.1 ps, and observed a different spectral dependence of the ΔR/R signal. This can be seen in Fig. 3a (blue curve labelled 'second irradiation'), where the minimum has now moved below 1200 nm and no longer matches the measurements previously obtained on the very same spot (Figs. 2a and 3a, red curves).

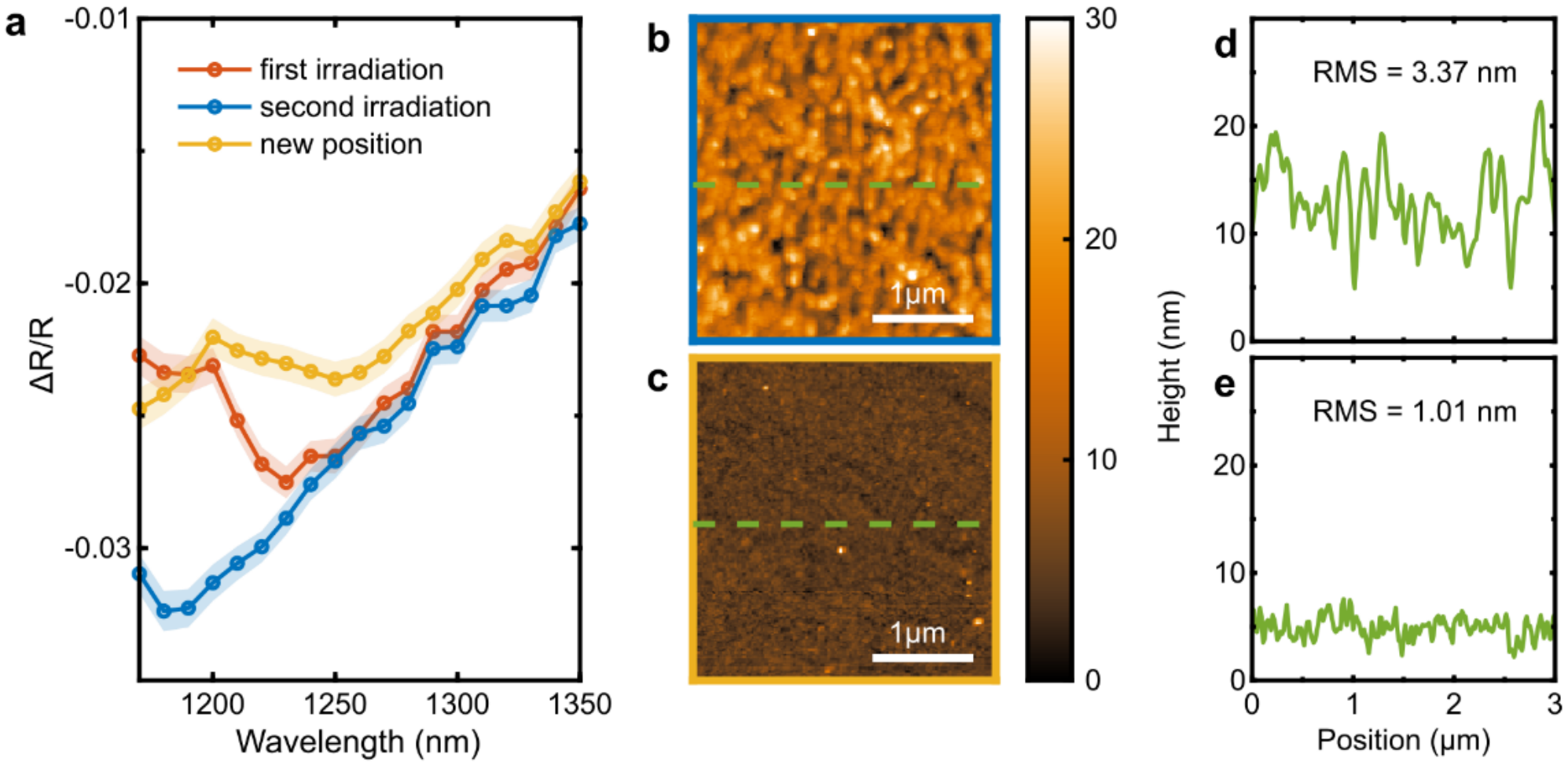

**Fig. 3. Effects of cumulative damage.** (a) Transient reflectance ΔR/R spectra recorded on the original location after the experiments reported in Fig. 2a (red and blue curves) and on an intact spot (yellow curve). AFM maps of the damaged (b) and intact (c) areas. Line profiles of the AFM maps for the damaged (d) and intact (e) areas, with RMS.

The presence of surface damage was confirmed by atomic force microscopy (AFM) on the irradiated area, as shown in Fig. 3b. The repetition of the TG measurement in another spot, using nominally the same parameters as before at a time delay of 0.1 ps (Fig. 3a, yellow curve), again reveals a minimum in the same spectral region, although both its intensity and position are slightly different. This small discrepancy is most likely due to the inhomogeneity of the multilayer structure (see Extended Data Fig. 1). AFM inspection after this single irradiation confirms that the newly illuminated area remains intact and free of defects (Fig.

3c). These results demonstrate that the observed damage originates from the prolonged exposure required during the exploratory phase of the experiment; once the measurement conditions were established, repeating the experiment on a fresh spot did not induce any detectable degradation, confirming both the robustness of the approach and the reliability of the measured response under controlled exposure conditions.

In summary, in this work we showed that an optically induced TG can act as an ultrafast, reconfigurable coupling element to excite BPP modes in HMMs. By comparing pump-probe spectra acquired with a single pump pulse (no interference pattern) and with two interfering pumps (TG-activated), we observe the emergence of a distinct spectral feature in the ΔR/R signal within the wavelength range expected for grating-coupled BPP excitations[13]. The absence of comparable features on the $Al_2O_3$/silica reference sample (see Extended Data Fig. 3) is consistent with the interpretation that the measured response is dominated by a BPP mode excited by the TG. Thus, the physical picture emerging from the measurements is compatible with TG-assisted phase matching: the spatiotemporal periodic permittivity modulation provides additional in-plane momentum that can couple the probe light to BPP modes supported by the HMM multilayer. Simulations enforce this conclusion by reproducing a BPP-like feature at a wavelength close to the experimentally observed one and by yielding near-field distributions consistent with the corresponding eigenmodes of the HMM. A key outcome of this study is the identification of the temporal window on the order of 1 ps over which the TG can function as a coupling element. This is consistent with a rapid decay of the photo-induced permittivity contrast due to carrier relaxation and diffusion[16], making the concept presented here interesting when targeting the manipulation of coherent phenomena occurring below 1 ps, such as polariton formation and collapse[21], and enabling ultrafast operations in quantum devices[22]. Overall, the present results suggest that TG writing can provide a route to spatiotemporally reconfigure coupling to otherwise momentum-mismatched BPP modes, offering an alternative to permanently nanostructured gratings and potentially enabling ultrafast control of optical modes excitation. Important next steps will be to enhance the effective coupling strength and the mode quality by optimizing the TG contrast and spatial profile (including its depth distribution), refining the multilayer design to reduce losses and sharpen BPP modes, exploring material platforms with more favorable photo-induced index dynamics, and adopting protocols that limit damage (e.g., single-shot or reduced-duty-cycle measurements). Such developments could broaden the accessible parameter space for ultrafast, programmable nanophotonics based on time-dependent permittivity landscapes in optical materials.

**References**

1. Lobet, M. *et al.* New Horizons in Near-Zero Refractive Index Photonics and Hyperbolic Metamaterials. *ACS Photonics* **10**, 3805–3820 (2023).
2. Lu, D. *et al.* Nanostructuring Multilayer Hyperbolic Metamaterials for Ultrafast and Bright Green InGaN Quantum Wells. *Adv. Mater.* **30**, 1706411 (2018).
3. Avrutsky, I., Salakhutdinov, I., Elser, J. & Podolskiy, V. Highly confined optical modes in nanoscale metal-dielectric multilayers. *Phys. Rev. B* **75**, 241402 (2007).
4. Valentine, J. *et al.* Three-dimensional optical metamaterial with a negative refractive index. *Nature* **455**, 376–379 (2008).
5. Teng, H., Chen, N., Hu, H., García De Abajo, F. J. & Dai, Q. Steering and cloaking of hyperbolic polaritons at deep-subwavelength scales. *Nat. Commun.* **15**, 4463 (2024).
6. Maccaferri, N. *et al.* Hyperbolic Meta-Antennas Enable Full Control of Scattering and Absorption of Light. *Nano Lett.* **19**, 1851–1859 (2019).

7. Kuttruff, J. *et al.* Magneto-Optical Activity in Nonmagnetic Hyperbolic Nanoparticles. *Phys. Rev. Lett.* **127**, 217402 (2021).
8. Suresh, S. *et al.* Enhanced Nonlinear Optical Responses of Layered Epsilon-near-Zero Metamaterials at Visible Frequencies. *ACS Photonics* **8**, 125–129 (2021).
9. Maccaferri, N. *et al.* Enhanced Nonlinear Emission from Single Multilayered Metal–Dielectric Nanocavities Resonating in the Near-Infrared. *ACS Photonics* **8**, 512–520 (2021).
10. Palermo, G. *et al.* Hyperbolic dispersion metasurfaces for molecular biosensing. *Nanophotonics* **10**, 295–314 (2020).
11. Liu, Z., Lee, H., Xiong, Y., Sun, C. & Zhang, X. Far-Field Optical Hyperlens Magnifying Sub-Diffraction-Limited Objects. *Science* **315**, 1686–1686 (2007).
12. Harwood, A. C. *et al.* Space-time optical diffraction from synthetic motion. *Nat. Commun.* **16**, 5147 (2025).
13. Maccaferri, N. *et al.* Designer Bloch plasmon polariton dispersion in grating-coupled hyperbolic metamaterials. *APL Photonics* **5**, 076109 (2020).
14. Eichler, H. J., Günter, P. & Pohl, D. W. *Laser-Induced Dynamic Gratings*. vol. 50 (Springer Berlin Heidelberg, Berlin, Heidelberg, 1986).
15. Bencivenga, F. *et al.* Four-wave mixing experiments with extreme ultraviolet transient gratings. *Nature* **520**, 205–208 (2015).
16. Pashina, O. *et al.* Excitation of surface plasmon-polaritons through optically induced ultrafast transient gratings. *Phys. Rev. Appl.* **25**, 014002 (2026).
17. Berté, R., Possmayer, T., Tittl, A., Menezes, L. D. S. & Maier, S. A. All-optical permittivity-asymmetric quasi-bound states in the continuum. *Light Sci. Appl.* **14**, 185 (2025).
18. Sivan, Y. & Spector, M. Ultrafast Dynamics of Optically Induced Heat Gratings in Metals. *ACS Photonics* **7**, 1271–1279 (2020).
19. Alber, L., Scalera, V., Unikandanunni, V., Schick, D. & Bonetti, S. NTMpy: An open source package for solving coupled parabolic differential equations in the framework of the three-temperature model. *Comput. Phys. Commun.* **265**, 107990 (2021).
20. COMSOL Multiphysics® www.comsol.com COMSOL AB, Stockholm, Sweden.
21. Kuttruff, J. *et al.* Sub-picosecond collapse of molecular polaritons to pure molecular transition in plasmonic photoswitch-nanoantennas. *Nat. Commun.* **14**, 3875 (2023).
22. Timmer, D. *et al.* Ultrafast transition from coherent to incoherent polariton nonlinearities in a hybrid 1L-WS2/plasmon structure. *Nat. Nanotechnol.* **21**, 216–222 (2026).
23. Capotondi, F. *et al.* Invited Article: Coherent imaging using seeded free-electron laser pulses with variable polarization: First results and research opportunities. *Rev. Sci. Instrum.* **84**, 051301 (2013).
24. Allaria, E. *et al.* Highly coherent and stable pulses from the FERMI seeded free-electron laser in the extreme ultraviolet. *Nat. Photonics* **6**, 699–704 (2012).
25. Mincigrucci, R. *et al.* Advances in instrumentation for FEL-based four-wave-mixing experiments. *Nucl. Instrum. Methods Phys. Res. Sect. Accel. Spectrometers Detect. Assoc. Equip.* **907**, 132–148 (2018).
26. Raimondi, L. *et al.* Microfocusing of the FERMI@Elettra FEL beam with a K–B active optics system: Spot size predictions by application of the WISE code. *Nucl. Instrum. Methods Phys. Res. Sect. Accel. Spectrometers Detect. Assoc. Equip.* **710**, 131–138 (2013).
27. Foglia, L. *et al.* Extreme ultraviolet transient gratings: A tool for nanoscale photoacoustics. *Photoacoustics* **29**, 100453 (2023).

28. Yakubovsky, D. I., Arsenin, A. V., Stebunov, Y. V., Fedyanin, D. Yu. & Volkov, V. S. Optical constants and structural properties of thin gold films. *Opt. Express* **25**, 25574 (2017).

**Methods**

Samples fabrication. The metal-insulator-metal HMM structure made of alternating layers of [Au (15 nm)/$Al_2O_3$ (30 nm)]x8 was prepared on fused silica substrates by electron beam evaporation in a custom-made vacuum chamber at the base pressure of $1 \times 10^{-6}$ mbar. Au and $Al_2O_3$ layers have been deposited at a rate of 0.2 and 0.4 Å $s^{-1}$, respectively. The layer thicknesses have been measured by quartz microbalance. An additional reference sample with 30 nm of $Al_2O_3$ on top of fused silica was also prepared using the same system and parameters. Before the deposition process, the substrates were cleaned in piranha solution (3:1 concentrated sulfuric acid to 30% hydrogen peroxide) for 15 minutes in order to remove possible organic residues.

Transient grating experiment. All the measurements have been carried out at the DiProI end-station[23] of the FERMI FEL source[24]. The mini-TIMER setup mounted inside the DiProI vacuum vessel allowed for the generation of the EUV TG[15,25]. Thanks to three carbon-coated mirrors, the FEL pulses are divided in two halves travelling along different optical paths, which are subsequently recombined at the sample plane, as schematically shown in Extended Data Fig. 2a. The position and orientation of the mirrors can be adjusted in order to ensure perfect spatial and temporal overlap at the sample plane. The $L_G$ spatial periodicity of the resulting TG is then given by

$$L_G = \frac{\lambda_{FEL}}{2 \sin \vartheta},$$

where $\lambda_{FEL}$ is the FEL wavelength, and $\theta$ is half of the crossing angle between the two FEL optical paths. In our experiment, $\lambda_{FEL} = 22.7$ nm (p-polarized), and $\vartheta = 1.7$ deg, corresponding to $L_G \approx 383$ nm. The spatial extent of the TG region on the sample plane was approx. 170 µm x 280 µm, as set by properly regulating the Kirkpatrick-Baez active optics system[26] located upstream the end-station. Considering the typical pulse energies used in the experiments, 6.7 µJ for the measurements shown in Figs. 2a and 3a and 4.3 µJ for the measurement shown in Fig. 2b, we can estimate an average fluence of 14 mJ/cm$^2$ and 9 mJ/cm$^2$, respectively.

To probe the reflectance of the HMM structure, we employed light in the near-infrared (NIR) range generated by a TOPAS-PRIME optical parametric amplifier (OPA) system, and hitting the sample at 45 deg of incidence. As shown in Extended Data Fig. 2b, the p-polarized 800 nm light pulses routinely available at the beamline were down-converted to the 1150 – 1550 nm range, with a spot size of approx. 180 µm x 150 µm (FWHM) at the sample plane. A dedicated delay line allowed us to probe the HMM response at a certain delay with respect to the TG generation. The duration of the FEL pulses was approx. 60 fs, while the duration of the probe pulses from the OPA was estimated to be approx. 70 fs. Thanks to a removable mirror on a magnetic mount, the optical transport system allowed us to deliver to the sample either 800 nm pulses, for diagnostic purposes, or the NIR light we used for probing the reflectance of the HMM structure. In this latter case, the light reflected by the sample was collected by a Ge photodetector (Thorlabs PDA50B2), and acquired via an oscilloscope triggered at the pulse repetition rate (50 Hz, for both FEL and NIR light). For each probe wavelength in the 1150 – 1550 nm range (sampled every 10 nm), we acquired 2000 single shot traces for both the reflected probe channel and the intensity monitor channel. Each trace was integrated over a fixed temporal window, yielding the integrated signals $I_1$ and $I_0$ for the reflected probe and intensity monitor channels, respectively. The reflectance-related signals were then calculated as $R = I_1/I_0$. This procedure was applied to both the TG measurement $R_{TG}$ and the corresponding reference measurement $R_{ref}$ acquired without TG excitation. Hence, the pump induced relative signal change was calculated as $\Delta R/R = (R_{TG} - R_{ref})/R_{ref}$. The uncertainties of $I_1$ and $I_0$ were estimated as $3\sigma/\sqrt{N}$, corresponding approximately to a 99.7% confidence interval under a normal approximation, and the uncertainties of $R_{TG}$, $R_{ref}$, and $\Delta R/R$ were obtained by standard error propagation.

Numerical simulations. The electromagnetic properties of the multilayer system have been verified by means of two-dimensional FEM simulations performed with COMSOL Multiphysics[20].
Initially, the absorbed power profile at the FEL pump wavelength and incidence angle was calculated by means of the TMM, as implemented in NTMpy[19]. The full $Al_2O_3$ (30 nm)/[Au (15 nm)/$Al_2O_3$ (30 nm)]x8/$SiO_2$ stack was considered, and the refractive indices were obtained from CXRO (https://henke.lbl.gov/optical_constants/). For a pump wavelength of 22.7 nm, we have that $n(Al_2O_3) = 0.910 + i0.069$, $n(Au) = 0.852 + i0.183$, $n(SiO_2) = 0.936 + i0.044$. Then, the resulting absorbed power profile was imported in COMSOL Multiphysics in order to describe the effect of the FEL-induced TG in a frequency-domain simulation in the NIR range. To this end, we made use of the "Electromagnetic Waves, Frequency Domain" physics interface available in the "Wave Optics" module, which solves for the spatial distribution of a time-harmonic electromagnetic field, assuming an $\exp(i\omega t)$ time dependence. Defining "z" as the out-of-plane coordinate along the multilayer stacking direction, and "x" as the coordinate in the sample plane along the direction of the TG modulation, we modelled the sample as a two-dimensional system confined in the xz-plane, since the extent of the grating in the "y" axis can be considered infinite with respect to both the grating period and the simulated region size. Indicating with $A(z)$ the normalized absorbed power profile, we introduced the effect of the TG as a spatial modulation of the refractive index $n(\lambda_{NIR}; x, z)$ in the simulation domain according to the following form[15,27]

$$n(\lambda_{NIR}; x, z) = n(\lambda_{NIR}) + i\Delta k A(z)\left[\frac{1 + \cos(k_G x)}{2}\right],$$

where $n(\lambda_{NIR})$ is the refractive index at the $\lambda_{NIR}$ probe wavelength, $\Delta k$ ($\in \mathbb{R}$) is the maximum value of the refractive index modulation, and $k_G$ is the amplitude of the TG wavevector as calculated according to $2\pi/L_G$. For the simulations reported in the main text, we selected a grating period $L_G = 385$ nm. For simplicity, the discussion in the main text refers to the experimental nominal value of 383 nm, since the difference is negligible for the present analysis. As material parameters, we considered $\Delta k = 5$, $n(Al_2O_3) = 1.70$ (constant in $\lambda_{NIR}$), $n(SiO_2) = 1.45$ (constant in $\lambda_{NIR}$), and $n(Au)$ as a function of $\lambda_{NIR}$ from Ref. [28] via https://refractiveindex.info/. To simulate the system response at several $\lambda_{NIR}$ values, we performed a parametric sweep in the range 1000 – 1600 nm, considering a p-polarized wave impinging on the multilayer stack at 45 deg. Along the z-axis, the simulation region was capped by perfectly matched layers for avoiding unwanted reflections at the external boundaries (vacuum on the top, semi-infinite $SiO_2$ substrate on the bottom), while we considered periodic boundary conditions along the x-axis. From the simulation results, we were able to extract both the $\Delta R/R$ signal and the spatial distribution of the electric field amplitude, as shown in Fig. 2. In particular, the simulated $\Delta R/R$ signal was obtained as $(R_{TG} - R_{ref})/R_{ref}$, where $R_{TG}$ is the reflectance obtained with the TG modulation described above, while $R_{ref}$ is the reflectance obtained with $\Delta k = 0$ (i.e., without modulation of the refractive index).

Finally, we calculated the near-field profile of the BPP eigenmode (Fig. 2e in main text) by applying the eigenfrequency solver implemented in COMSOL Multiphysics on the full unmodulated multilayer stack including the substrate.

Atomic force microscopy imaging. AFM measurements were performed using a Nanosurf FlexAFM operating in tapping mode, under ambient pressure and temperature conditions. Commercial silicon cantilevers (Tap190Al-G, from BudgetSensors), with nominal spring constant of 48 N/m and resonance frequency of 190 kHz, were used. For acquiring the images, we selected a scan size of 8 x 8 $\mu m^2$, an image size of 512 x 512 $px^2$, and a scan rate of 0.5 Hz. Topography data were obtained from the Z-piezo signal. Image processing was carried out using Nanosurf Naio, applying a liner slope removal to correct for sample tilt. Quantitative analysis was performed by extracting a single line cut (Fig. 3b and c, dashed lines), and calculating the root mean square (RMS) value of the height variations.

**Acknowledgements**
This work was funded by the Swedish Research Council (Grants No. 2021-05784 and No. 2025-04734), the Knut and Alice Wallenberg Foundation (Grant No. 2023.0089), and the European Research Council (ERC Starting Grant No. 101116253 'MagneticTWIST'). PV acknowledges support from the Spanish Ministry of Science and Innovation under the Maria de Maeztu Units of Excellence Programme CEX2020-001038-M and the Project PID2024-155776NB-I00 (MICINN/FEDER). DG and RK acknowledge funding from the European Union under the Horizon 2020 Program, FET-Open: DNA-FAIRYLIGHTS (Grant No. 964995) and the European Union Program HORIZON-Pathfinder-Open: 3D-BRICKS (Grant No. 101099125). We acknowledge Stefano Bonetti for providing access to the workstation running COMSOL Multiphysics.

**Author contributions**
NM conceived the study and supervised the project. MP, LF, EP, AV, RM, FC, FB, DG, RK, TT and NM performed the pump-probe measurements. IN, MP, FC, EP and MBD developed the pump-probe experimental setup. MP developed the theory with input from NM and FB. DG and RK performed the sample fabrication. RT performed the AFM measurements. TT, MP, HK, PV and NM analyzed and discussed the results with input from LF, FB and FC. NM, TT, HK and MP wrote the manuscript with input from all the authors. TT, HK and MP contributed equally to this work.

**Competing interests**
The authors declare no competing interests.

## Extended Data

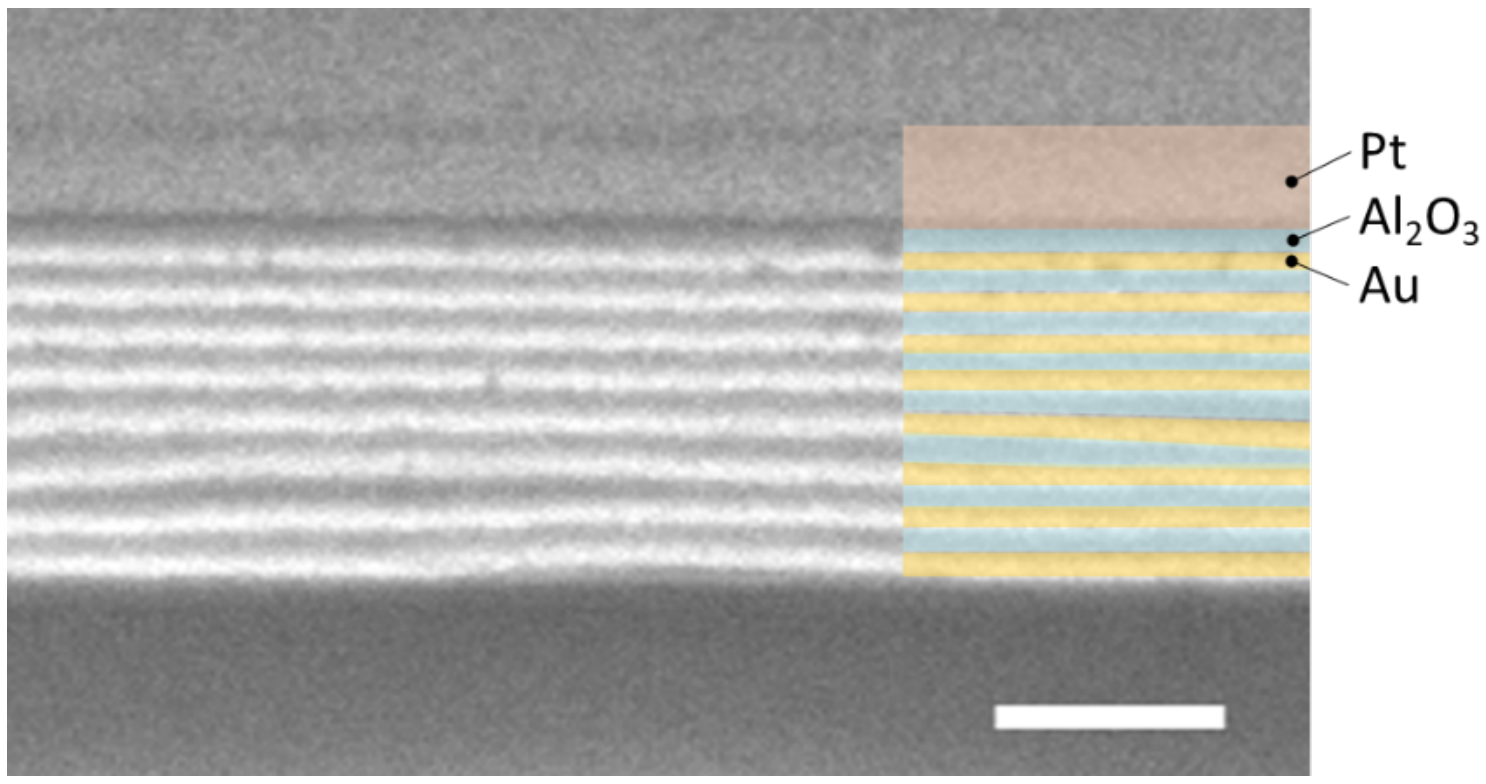

**Extended Data Fig. 1 | FIB-SEM cross section of the Au/$Al_2O_3$ hyperbolic metamaterial.** The alternating bright and dark contrast bands correspond to eight repeated metal dielectric Au/$Al_2O_3$ stacks. The Pt layer was deposited on top of the 30nm $Al_2O_3$ layer to take the cross section of the HMM structure. Scale bar: 200 nm.

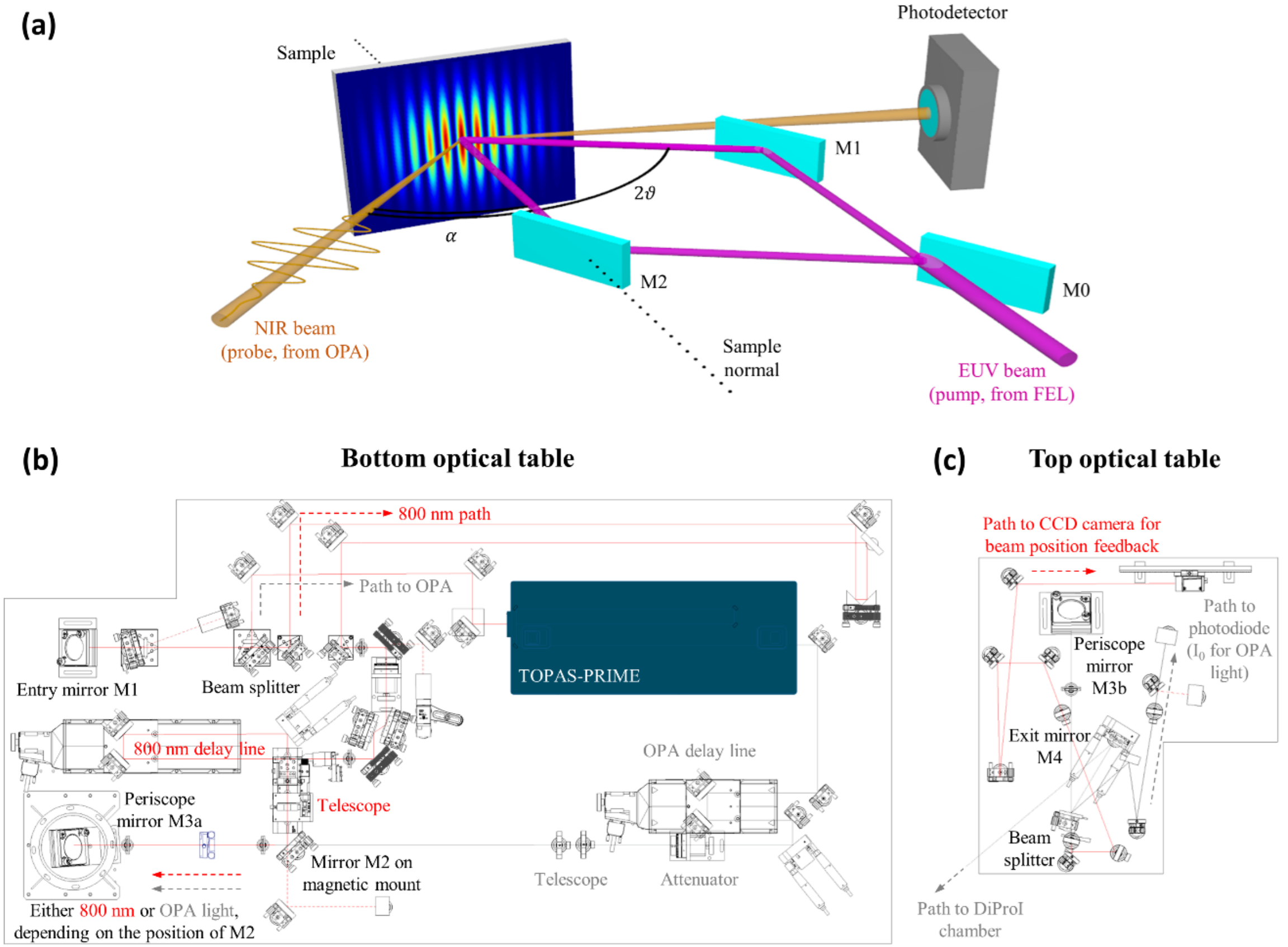

**Extended Data Fig. 2 | Transient grating pump-probe experimental setup.** The TG pump-probe experiment was performed at the DiProI end-station of the FERMI free-electron laser facility. (a) Schematic of the optical layout used for EUV TG pump-probe measurements. The EUV pulses from the FERMI free-electron laser are split into two pump beams and recombined at the sample plane under a total crossing angle $2\vartheta = 3.4$ deg, generating a transient intensity grating. A time delayed NIR probe beam from the OPA impinges on the sample at an angle $\alpha = 45$ deg, and the reflected signal is collected by a Ge photodetector. (b) Optical layout of the bottom optical table available at the DiProI end-station, showing the 800 nm paths, and the TOPAS-PRIME OPA. Thanks to a beam splitter, part of the 800 nm light doesn't reach the OPA, and it can be transported to the sample for diagnostic purposes. (c) Optical layout of the top optical table, showing the final beam delivery towards the DiProI vacuum chamber, together with the paths used for shot-by-shot beam position feedback and OPA intensity monitoring ($I_0$).

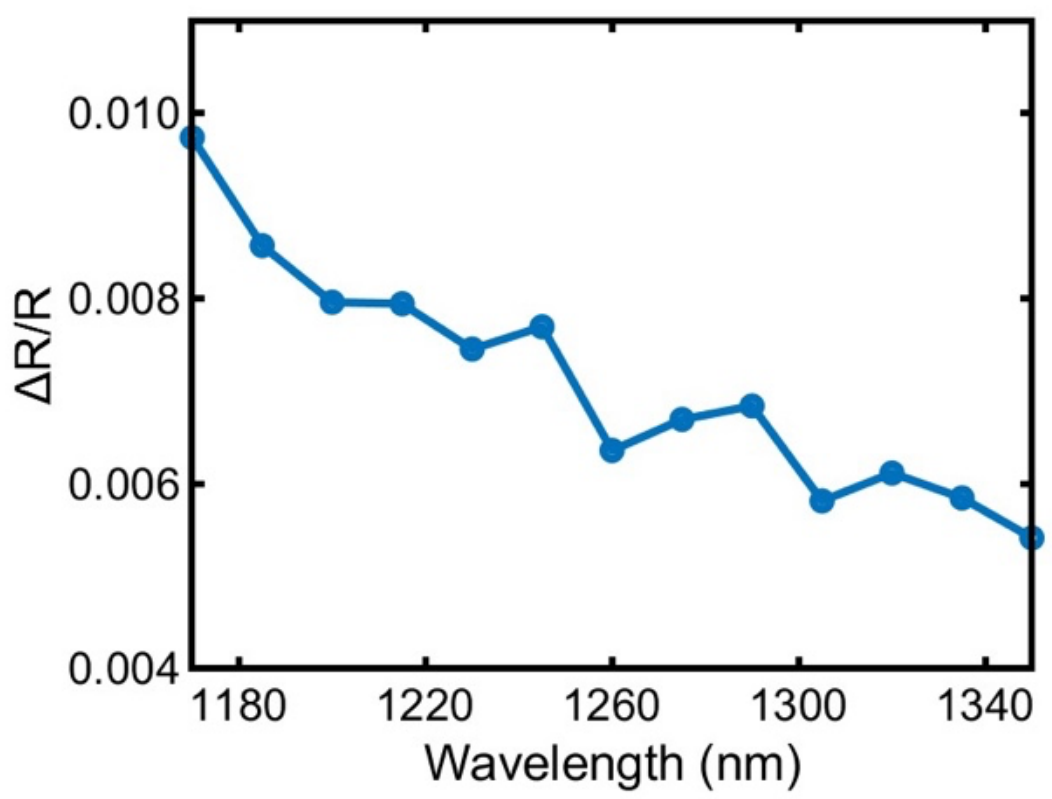

**Extended Data Fig. 3 | Reference $Al_2O_3/SiO_2$ control experiment.** Transient reflectance ΔR/R spectrum measured on the reference sample consisting of a 30 nm $Al_2O_3$ film deposited on a fused silica substrate under TG excitation. No pronounced spectral dip is observed in the investigated near-infrared wavelength range, in contrast to the HMM sample (see main text). This confirms that the spectral feature observed in the HMM does not arise from the $Al_2O_3$ layer alone, but it requires the $Au/Al_2O_3$ multilayer structure and can be assigned to the excitation of a BPP mode.